\documentclass[nopreprintline, 12pt, 
]{elsarticle}
\usepackage[utf8]{inputenc}
\usepackage{graphicx}
\usepackage{amsmath}
\usepackage[hyphens]{url}
\usepackage{listings}
\usepackage{booktabs}
\usepackage{multirow}
\usepackage{multicol}
\usepackage{tabulary}
\usepackage{tcolorbox}
\usepackage{xcolor}

\newcommand{\Qone}{RQ 1: How many games delay their release?}
\newcommand{\Qtwo}{RQ 2: What types of games delay release?}
\newcommand{\Qthree}{RQ 3: How is delaying a release perceived by players?}
\newcommand{\Qfour}{RQ 4: How likely are games with a non-concrete release date to be delayed?}
\graphicspath{ {./images/} }

\journal{Entertainment Computing} 

\begin{document}

\begin{frontmatter}

\title{An Empirical Study of Delayed Games on Steam}
\author[1]{Balreet Grewal}\corref{cor1}
\ead{balreet@ualberta.ca}
\author[2]{Dayi Lin} 
\ead{heylindayi@gmail.com}
\author[3]{Lars Doucet}
\ead{lars.doucet@gmail.com}
\author[1]{Cor-Paul Bezemer}
\ead{bezemer@ualberta.ca}
\cortext[cor1]{Corresponding author}
\affiliation[1]{organization={University of Alberta, Analytics of Software, Games and Repository Data (ASGAARD) lab},
            country={Canada}}

\begin{abstract}
The gaming industry is rapidly expanding. With over 2.7 billion players worldwide, game development has become increasingly challenging. To meet the ever-changing demands and expectations of players and due to unforeseen hindrances in the development process, game developers may require to delay the release of their game. We conducted an empirical study of 23,485 games on the Steam platform to analyze how often, and which games delayed their release date. We find that delaying a release is common: 48\% of the studied games had a delayed initial release. Games delayed their release by a median of 14 days. Games for which a release date range (e.g., ``Q1 2019") was specified, rather than a specific date were more likely to release within that range. Across different game genres, the percentage of games that delay release is similar (ranging from 48\% to 52\%). Finally, games with a delayed release are rated lower than games that release on time, but the difference is negligible. 
\end{abstract}

\begin{highlights}
\item The first large-scale study of Steam games that had delayed release dates.
\item Release date ranges are more likely to be met than specific release dates for games unless the release date range specifies the quarter and year. 
\item Games push release more often than not.
\item The portion of games that delay release is increasing.
\item Games with a delayed release are rated lower than games that release on time, but the difference is negligible.
\end{highlights}

\end{frontmatter}

\section{Introduction}
\label{section: Intro}

In early 2021, the ever-growing gaming industry has reached a value of over US \$300 billion and has over 2.7 billion players worldwide~\cite{accenture}. Within this growing market, creating a successful game has many challenges, including, but not limited to,  hindrances that occur during the development phase. For multiple reasons, developers may be inclined to change the initial release date of a game under development to ensure that the final product is as planned. 

In this paper, we study how often games have a delayed initial release. Prior studies have explored the phenomenon of delayed releases in the movie industry~\cite{Einav2009, EinavRelTiming}. Another has directed attention towards how developers select a release date~\cite{Engel2018}. However, to the best of our knowledge, our study is the first to examine delayed releases for games.

In this paper we present a study of 23,485 games on the Steam platform, the largest online game distribution platform. Our study first provides insights about how often games have a delayed release. Second, we study what types of games delay their release, and whether there are differences in user-perceived quality between games that delay their release, and those that release on time.  
In particular, we focus on the following four research questions: 
\begin{quote}
    \textbf{\Qone} We begin by analyzing how often games miss their initial release dates. We found that games delayed release a median of 1 time and a median of 14 days. Further, an increasing portion of games (from 28\% in 2016 to 54\% in 2020) delayed release each year.
    
    \textbf{\Qtwo} We investigate whether certain genres of games and game tags delay more often than others. The portion of early access games that delayed their release more than once was larger than the portion of non-early access games. Similarly, the portion of indie games that delayed release more than once was larger than the portion of non-indie games. Further, we found that the percentage of games that delayed release across different game genres ranged from 48\% to 52\%.
    
    \textbf{\Qthree} We examine the positive review rates for different genres of games to garner how players perceive delaying a release in each genre. We found that the median value of the positive review rate of games with a delayed release was lower than games that were on-time, but the difference was negligible.  
    
    \textbf{\Qfour} We compare games that have a non-concrete release date (e.g., ``Q1 2020" or ``soon") with games that have a specific (concrete) release date. We found that games with non-concrete release dates released more often on time than games with a concrete release date, with the exception of games that specify a year and quarter of release. Games with a non-concrete release date range posted their non-concrete release date before entering the release date range period, and ambiguous non-concrete release dates (such as ``soon" or ``TBA") released within 3 months. 
\end{quote}

Our study shows that delaying a game's release is common. In addition, games with a delayed release are rated lower than games that are released on time, but the difference is negligible for most types of games. Game developers can leverage this knowledge when deciding whether they should delay the release of their games.

The remainder of this paper is organized as follows. Section~\ref{section:background} provides background information on the release dates on Steam and discusses related work. Section~\ref{section:method} presents our methodology for the study, followed by Sections~\ref{section:RQ1} through~\ref{section:RQ4} which present our results. Section~\ref{section:Discuss} presents a discussion of our findings, and Section~\ref{section:threat} discusses the threats to validity of our study. Finally, Section~\ref{section:conclude} concludes our paper.

\section{Background}
\label{section:background}

In this section, we provide a background into release dates on Steam and prior related research about release dates and mining game distribution platforms.

\subsection{Release dates on Steam}

Steam\footnote{\url{https://store.steampowered.com}}, developed by Valve Corporation, is a digital game distribution platform. Steam is currently the largest digital distribution platform for gaming with over 45,000 games~\cite{Steamspy} and roughly 22 million concurrent players a day~\cite{SteamDB}. The Steam platform consists of two major components for users: the Steam Store~\cite{SteamStore} and the Steam Community~\cite{SteamCommunity}. For developers, Steamworks~\cite{Steamworks} is vital in assisting the process of releasing a game on the Steam platform.

Steamworks~\cite{Steamworks} is a set of tools for developers, created by Valve, to manage games and the associated Steam Store for the game. For developers to release a game in the Steam Store, they must first provide company identification, payment information, and tax information. While a game is under development, developers can choose to set up a `Coming Soon' page to collect feedback and allow for users to wishlist the game.  Prior to August 2019, developers could change the release date of their game multiple times whenever desired. Some developers would intentionally delay the release date of their game in order to remain on certain lists on Steam (e.g., `Upcoming games') to gain attention from players~\cite{SteamRelChange}. To ensure that games that are in-fact releasing soon are seen by players on the Steam platform, developers must now contact Valve to change the release date of their game~\cite{Steamworks, PCSteamRel}.

To set a release date on a game's Steam Store page, developers can post a specific date (e.g., April 23, 2018) or a custom string (e.g., `Coming Early 2020'). If a custom string is displayed for a game's release date, developers must still have a specific release date set at the same time. Once a game is released on Steam, the specific release date will be displayed on the Steam Store page for the game.

\subsection{Release Dates}
Prior work on changes in release dates focuses on the movie industry. Einav and Ravid~\cite{Einav2009} studied the stock market's response to a change in a movie's opening date. They conclude that a change in opening dates in movies is perceived negatively in the market. Adding on, Einav~\cite{EinavRelTiming} studied the release date timing of movies and concludes that releases of movies are clustered together, especially around holidays. 

Engelst{\"a}tter and Ward.~\cite{Engel2018} studied how a video game publisher's choice of release date is affected by various conditions such as competition, and genre. The study presents that delaying the release of a game to avoid competition in similar game genres can lead to an increase in profits. Our study expands upon these findings to investigate the implications of delaying the release date of a game for developers and the perception of games that delay release.

\subsection{Mining Game Distribution Platforms}
There are a plethora of papers that have mined data from online game distribution platforms~\cite{BlackburnCheatingGames, ChambersOnlineGames, SifaCrossGames, YannakakisAI, WANG2020100338, BAILEY2019100299} for various purposes. Notably, Foxman et al.~\cite{FoxmanVR} analyzed common game genres among virtual reality (VR) games and their user ratings by crawling for the number of owners and ratings from a third-party API and the Steam platform. Similarly, Epp et al.~\cite{rain2021vr} gathered information such as price history and reviews from Steam and third-party websites to study the trends from VR games and their associated reviews. Li and Zhang~\cite{LiGameTags} crawled Steam to present a new approach to video game genre classification and to analyze connections between different game genres. Lin et al. have also implemented data mining approaches to analyze user reviews~\cite{Lin2018reviews}, early access games~\cite{Lin16eag}, urgent updates~\cite{Lin16urgent}, and bugs in computer games~\cite{Lin2019videos}. Similarly, Vu and Bezemer present an approach to improve the discoverability of indie games~\cite{Quang21}, and Lee et al. used customized crawlers to extract data  to study game modifications (mods)~\cite{Daniel2019nexusmods} and Minecraft mods~\cite{Lee2020curseforge}. In this paper, we also focus on extracting game data, game tags, and user review rates from Steam and third-party website APIs. However, we are the first to study the delay of game release dates.

Many studies have focused on mining and analyzing reviews from mobile app stores~\cite{ZhuGamePlay, TONG2021100077}. Vasa et al.~\cite{VasaMobileApp} studied user reviews on mobile apps and found that negative reviews are longer in length. Pagano and Maalej~\cite{PagnoApp} analyzed user reviews from the Apple App Store to inspect the content, timing, and impact of reviews. Fu et al.~\cite{FuWisCom} proposed an integrated system, WisCom, to analyze mobile app user reviews from a comment centric, app centric, and market centric view. Hassan et al.~\cite{safwat16replies} analyzed the dialogue between users and developers on the Google Play Store. They have also studied complaints from bad updates~\cite{safwat_tse} of top free-to-download apps on the Google Play Store and how developers recover following a bad update. Hu et al.~\cite{hu17hybrid} studied the consistency of rating and reviews of hybrid apps both available on Android and iOS. Our study analyzes the user review ratings for games that delay release and games that are on-time.

\section{Methodology}
\label{section:method}

\begin{figure}[t!]
\centering
\includegraphics[width=0.75\textwidth]{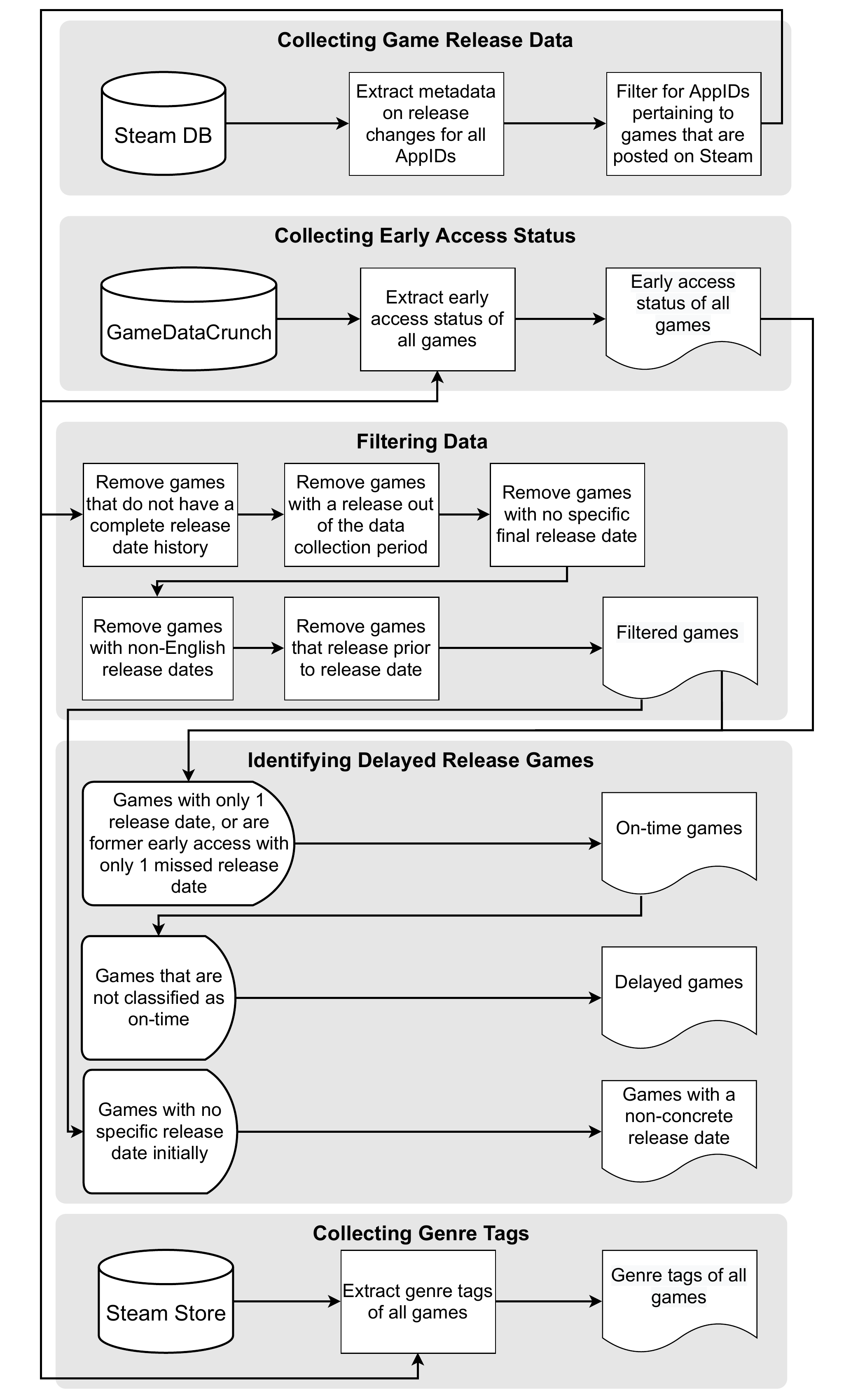} 
\caption{Overview of data collection.}
\label{fig:overview}
\end{figure}

In this section we cover our methodology for this study. The overview of the data collection and filtering process can be seen in Figure~\ref{fig:overview}. Table~\ref{table:summary} shows a brief description of the collected and filtered data set. The remainder of this section will highlight the data collection and filtering process in greater detail.

\subsection{Collecting game release data}
The data of the changes to a game's release date was kindly provided to us by the owners of the SteamDB website\footnote{\url{https://steamdb.info}}. SteamDB is a third-party service that tracks updates and changes for games on Steam. For each game release, we obtained the associated application ID (AppID) and release date history. The release date history can consist of a single or multiple entries regarding the release date changes of a game. In particular, each entry in the release date history consists of the type of release date change, the time of the release date change, and the old and new release dates (if applicable). There are three possible types of changes to a release date:
\begin{itemize}
    \item \textbf{Announcing a release date}: a release date has been added to the Steam store page of the game. 
    \item \textbf{Removing a release date}: a release date has been removed from the Steam store page of the game. 
    \item \textbf{Modifying a release date}: a release date has been changed on the Steam store page of the game. In this case there would be both an old and a new release date.
\end{itemize}
To ensure we only include games, and not other applications found on Steam (e.g., photo editing software, additional game content, game demos), we removed data for AppIDs that are not games. Our period of data collection ranged from January 1, 2016 until April 1, 2020. We also removed any games that ended up not being released on Steam.

\begin{table}[t]
\centering
\caption{Summary of collected data}
\label{table:summary}
\begin{tabular}{l r} 
 \toprule
 \# of unique AppIDs &  67,842 \\
 \# of game AppIDs & 34,230 \\
 \# of filtered games & 23,485 \\
 \# of delayed games & 9,262 \\
 \# of on-time games & 10,159 \\
 \# of games with a non-concrete release date & 4,064 \\
 \bottomrule
\end{tabular}
\end{table}

We use the release date history for games as a basis to categorize the games as delayed and on-time. A delayed game is one that delayed its release date, and an on-time game never delayed its release. We explain the categorization in detail further in this section.

\subsection{Collecting early access status}
To accurately sort games that  are on-time or delay release, we created a custom crawler to extract the early access status of games from the  GameDataCrunch\footnote{\url{https://www.gamedatacrunch.com}} website API. The methodology that was followed to collect the data by GameDataCrunch is described on the Fortress of Doors\footnote{\url{https://www.fortressofdoors.com/we-dont-actually-know-when-many-steam-games-were-released/}} blog. We performed the data collection on April 9, 2021. There are four early access statuses a game can occupy: (1)~unknown early access status, (2)~former early access, (3)~never early access, and (4)~current early access. We assumed games with an unknown early access status to not have been an early access game.

\begin{table}[t]
\centering
\footnotesize
\caption{Summary of the studied types of game. \# delayed is the number of games with a delayed release, \# on-time is the number of games with an on-time release, and \# non-concrete is the number of games with an initial non-concrete release date.\vspace{2mm}}
\label{table:Genres}
\begin{tabulary}{\textwidth}{L R R R R}
 \toprule
 \textbf{Genre} & \textbf{\# Games} & \textbf{\# Delayed } & \textbf{\# On-time } & \textbf{\# Non-concrete}\\
 \midrule
 Action & 8,798 & 3,375 & 3,678 & 1,745\\
 Adventure & 8,069 & 3,147 & 3,368 & 1,554\\
 Casual & 8,676 & 3,727 & 3,804 & 1,145\\
 Simulation & 3,721 & 1,594 & 1,461 & 666\\
 Strategy & 4,195 & 1,732 & 1,650 & 813\\
 RPG & 3,725 & 1,431 & 1,489 & 805\\
 \midrule
 \textbf{Game type} & \textbf{\# Games} & \textbf{\# Delayed} & \textbf{\# On-time} & \textbf{\# Non-concrete}\\
 \midrule
 Indie & 16,064 & 6,373 & 6,695 & 2,996\\
 Non-indie & 7,421 & 2,889 & 3,464 & 1,068\\
 \midrule
 Early access & 4,958 & 2,148 & 1,665 & 1,145\\
 Non-early access & 18,527 & 7,114 & 8,494 & 2,919\\
 \bottomrule
\end{tabulary}
\end{table}

\subsection{Filtering data}
First we removed games that do not have a complete release date history. A game has a complete release date history if it initially creates a release date on the Steam Store page and does not end with a removed release date change. Hence, we can analyze how many times a game changes release dates, and by how many days a release was delayed.

We then proceeded to remove games that have a final release date before or after the collection periods, as we cannot be certain if such games have a complete release history. Similarly, games that did not have a specific date for their final release were also removed. To accurately analyze games with a non-concrete release date, we removed games that have a release date that contains information in another language than English. 

Some games would release earlier than anticipated when comparing the initial release date and the actual release date of a game in the release date history. We removed games that released prior to their initial release date to avoid incorrectly categorizing games, as we noticed that many of such games were using an unrealistic initial release date (e.g., it was far in the future).

\subsection{Identifying delayed games}
To categorize a game as delayed or on-time we consider the release date history for an AppID and the set release date(s). To effectively read the release dates, we converted all of the `new release date' values from a Unix Epoch timestamp to a datetime value. For games with more than one release date change, we calculated the number of times a game delays release by counting the number of release date modifications and creations and then subtracting one (as the first created release date does not indicate a delay).

We then considered the following games to be on-time:
\begin{itemize}
    \item Games with only one entry in the release date history.
    \item Games with multiple entries in the release date history, that end up releasing on the date they initially specified 
    \item Games that release on the day when they posted the release date
    \item Games that delay once but were also former early access games. Early access games always have two release dates: one for the early access release, and one for the final release~\cite{Lin16eag}.
\end{itemize}
A game is categorized as a delayed game otherwise. We analyzed games that have a non-concrete release date separately.

\subsection{Collecting genre tags}
We used a custom crawler on the Steam platform to extract the game genre tags for each AppID. Table~\ref{table:Genres} displays the game genres we analyze. We selected the six most frequently used genre tags from the Steam Store~\cite{SteamStore} to examine. This selection ensures that we analyze game genres that are popular on Steam and that we have a large number (at least around 3,000) games for each category. We also studied early access games and indie games, as our expectation was that the early access model or indie development could affect the delay behaviour of games. We identified indie games using the `indie' game genre. We used the early access status as mentioned previously to extract the games that followed the early access model.

\section{\Qone}
\label{section:RQ1}

\textit{Motivation:} 
Before analyzing the characteristics of games that delay their release, we examine the commonality of a game delaying its initial release. Thus, we establish an understanding of the trends of delayed and on-time games for our following research questions. 

\textit{Approach:} 
For this research question we filtered out games with a non-concrete release date so that we may appropriately analyze the time taken for a game to release and effectively compare against games that release on time. To observe the number of days a game delays for, we calculate the number of days between the initially planned release date and the actual release date of the game. 

\begin{figure}[t]
\centering
\includegraphics[width=0.8\textwidth]{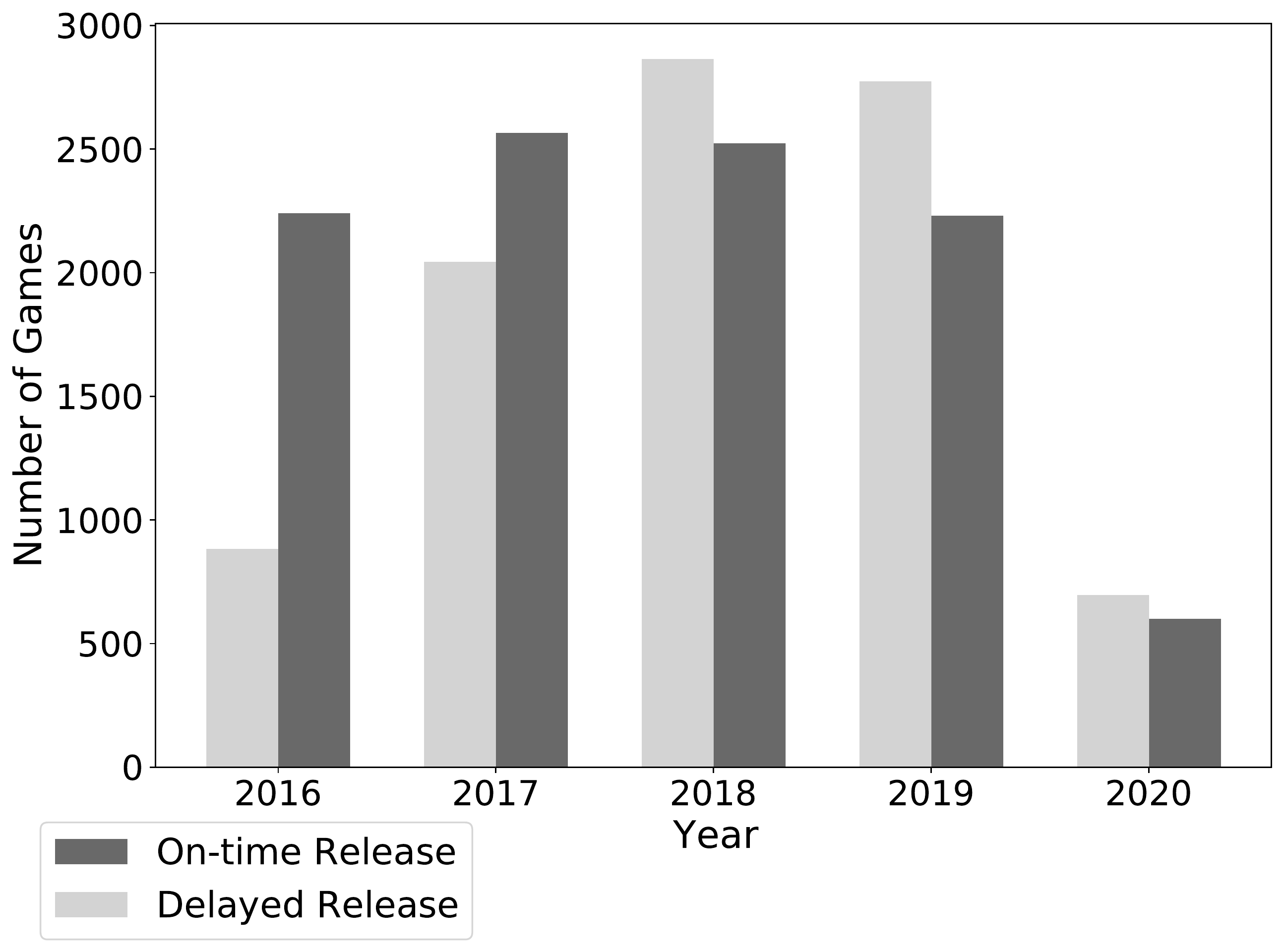}
\caption{Comparison of the number of games that release on-time and games that were delayed. Note that we count a game in the year in which it was released. Also note that the data for 2020 is partial (up to April 1).}
\label{fig:Yearly}
\end{figure}

\begin{figure}[t]
\centering
\includegraphics[width=0.8\textwidth]{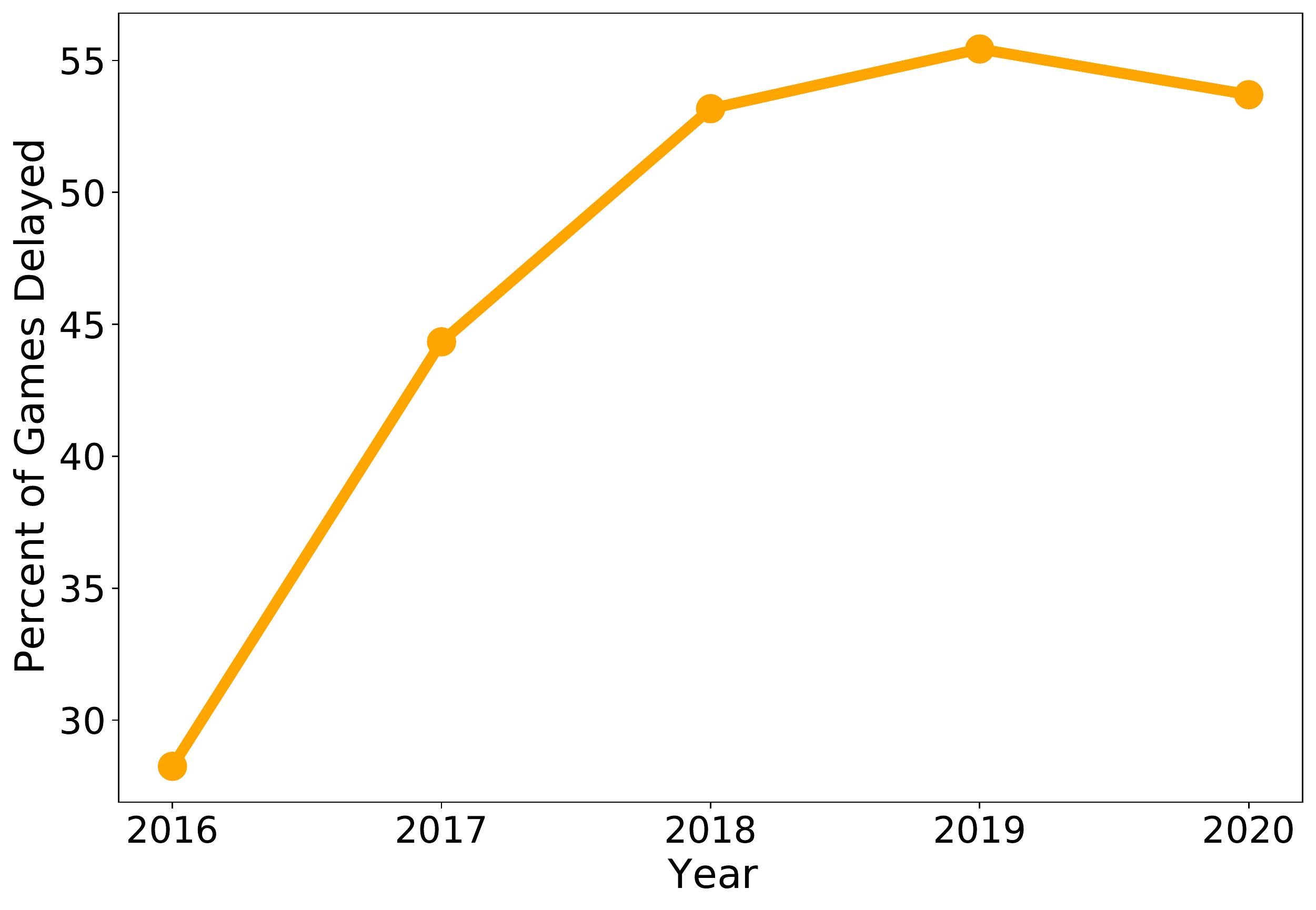}
\caption{ Plot of the percentages of games that delay their release each year. Note that we count a game in the year in which it was released. Also note that the data for 2020 is partial (up to April 1).}
\label{fig:PercentDelayed}
\end{figure}

\textit{Findings:} 
\textbf{The portion of games that delay their release increases every year.} As seen in Figure~\ref{fig:Yearly} and Figure~\ref{fig:PercentDelayed}, the portion of games that delayed their release increased from 28\% (882 games) in 2016 to 53\%-55\% (approximately 3000 games) in 2018,2019, and 2020. Thus, its has become increasingly common for a game to delay its release.

\begin{figure}[t]
\centering
\includegraphics[width=0.8\textwidth]{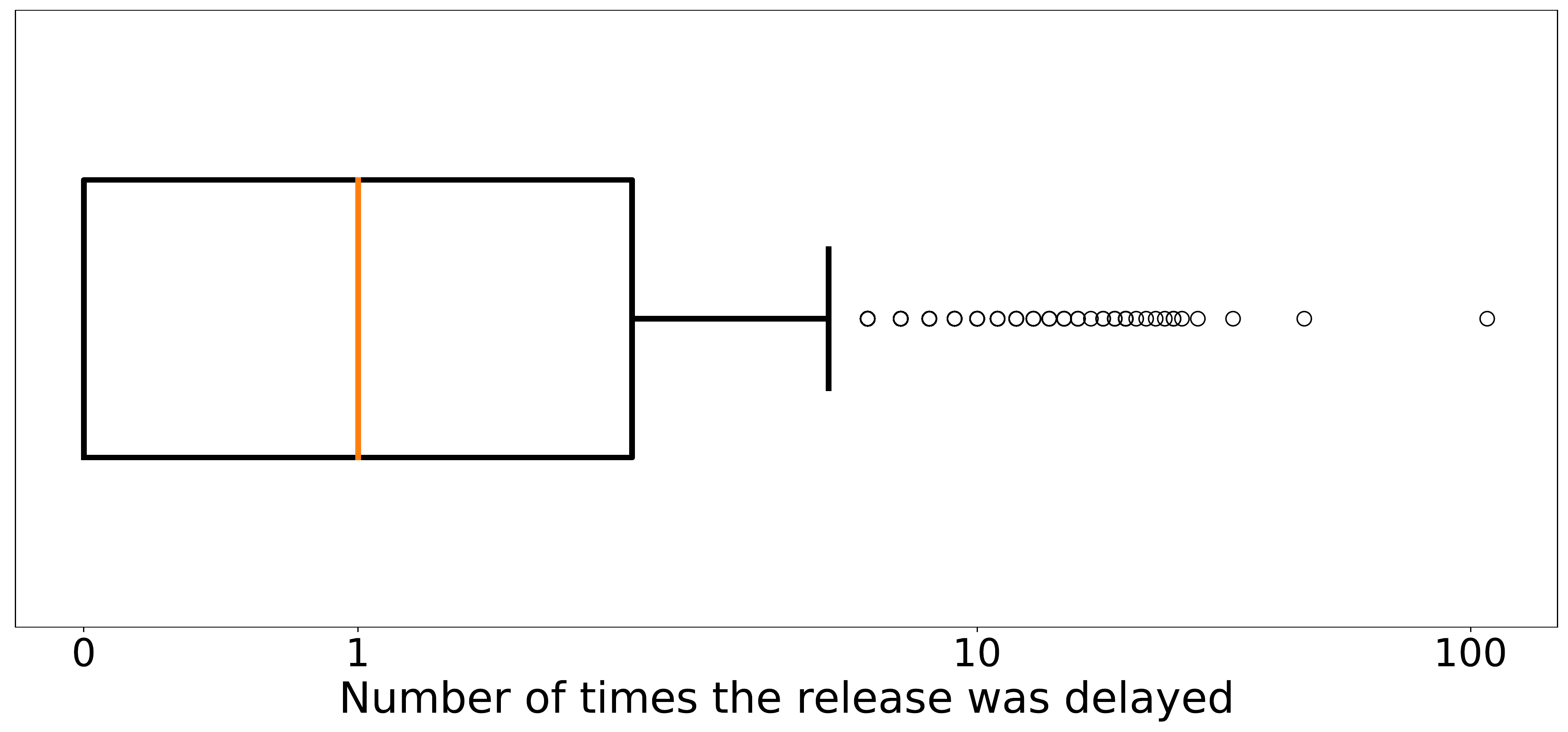}
\caption{Boxplot of the number of times a game delayed its release.}
\label{fig:Slips}
\end{figure}

\begin{figure}[t]
\centering
\includegraphics[width=0.8\textwidth]{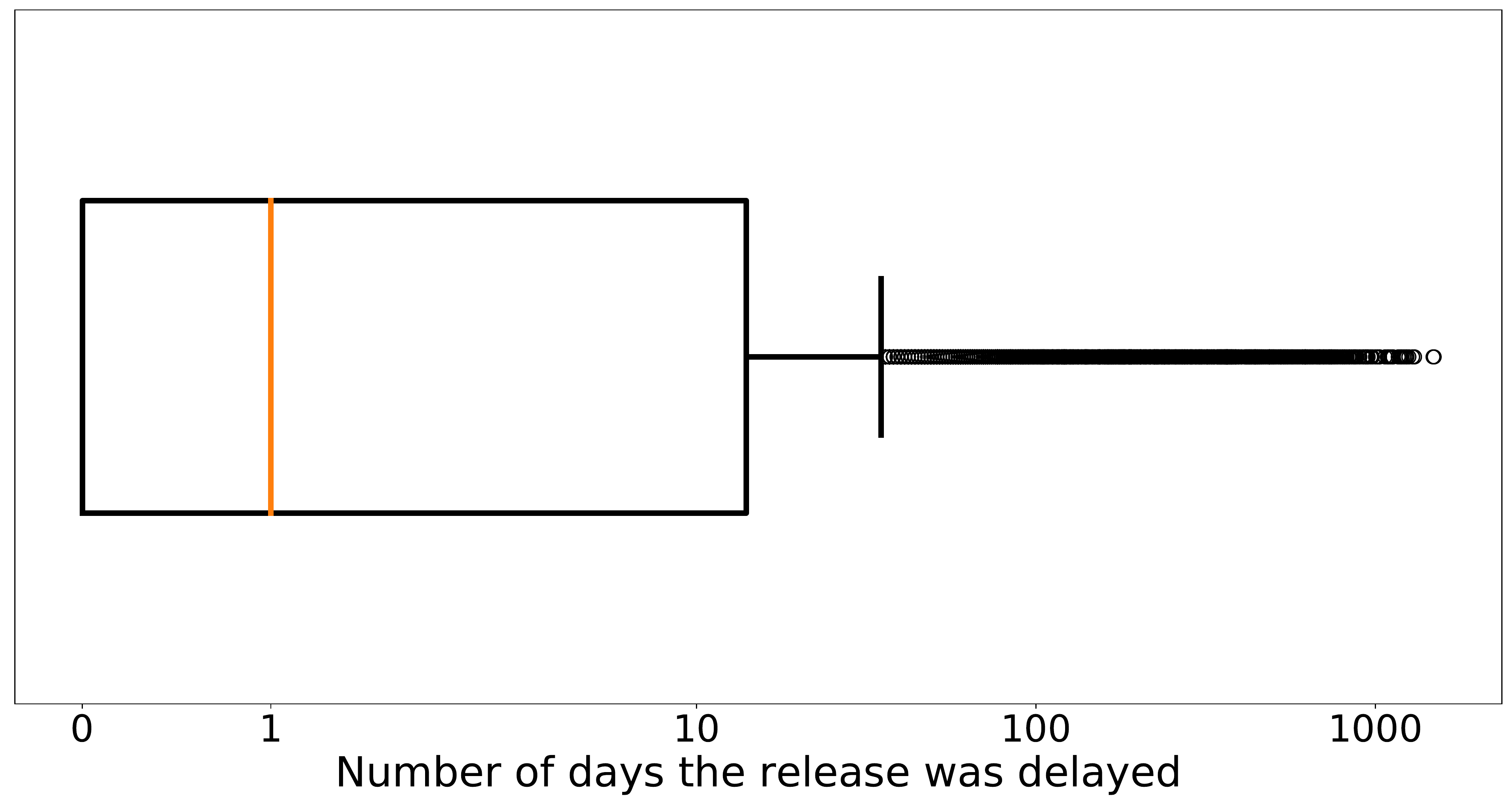}
\caption{Boxplot of the number of days a game was delayed by. Games that were on-time delayed for 0 days and games that delayed release did so for a median of 14 days.}
\label{fig:Days}
\end{figure}

\textbf{When a game delayed its release, it was delayed a median of 1 time.} Figure~\ref{fig:Slips} shows the boxplot pertaining to the number of times a game delays its initial release. There are 149 games that delay release at least 10 times. These games all released prior to August 2019 (before Valve required developers to have their release dates approved - see Section~\ref{section:background}). The large number of changes in the release date can potentially be attributed to developers changing their game's release date multiple times to stay on certain pages on the Steam store before release. 

\textbf{Games that delayed release were delayed by a median of 14 days}. Figure~\ref{fig:Days} shows the distribution of the number of days a game is delayed by. If we ignore the games that were on-time, the median delay was 14 days. There are 1,546 games that delayed their release for longer than three months; these games delayed a median of three times.  

\begin{tcolorbox}[left=2pt, right=2pt, top=2pt, bottom=2pt]
  \textit{\textbf{Summary:} Delaying the release of a game has become more common, with most games delaying once for a median of 14 days.}
\end{tcolorbox}

\section{\Qtwo}
\label{section:RQ2}

\textit{Motivation:} As there are games of multiple genres and types on Steam, their behaviour of delaying a release may be different. Thus, it is important to examine if games of a particular genre or type are more likely to delay their release.

\textit{Approach:} For this RQ, we studied the following types of games: (1)~games from different genres, (2)~early access vs. non-early access games, and (3)~indie vs. non-indie games. We collected the studied game genre tags and game types as mentioned in Section~\ref{section:method}. We again removed games with non-concrete release dates.

\begin{figure}[t!]
\centering
\includegraphics[width=0.8\textwidth]{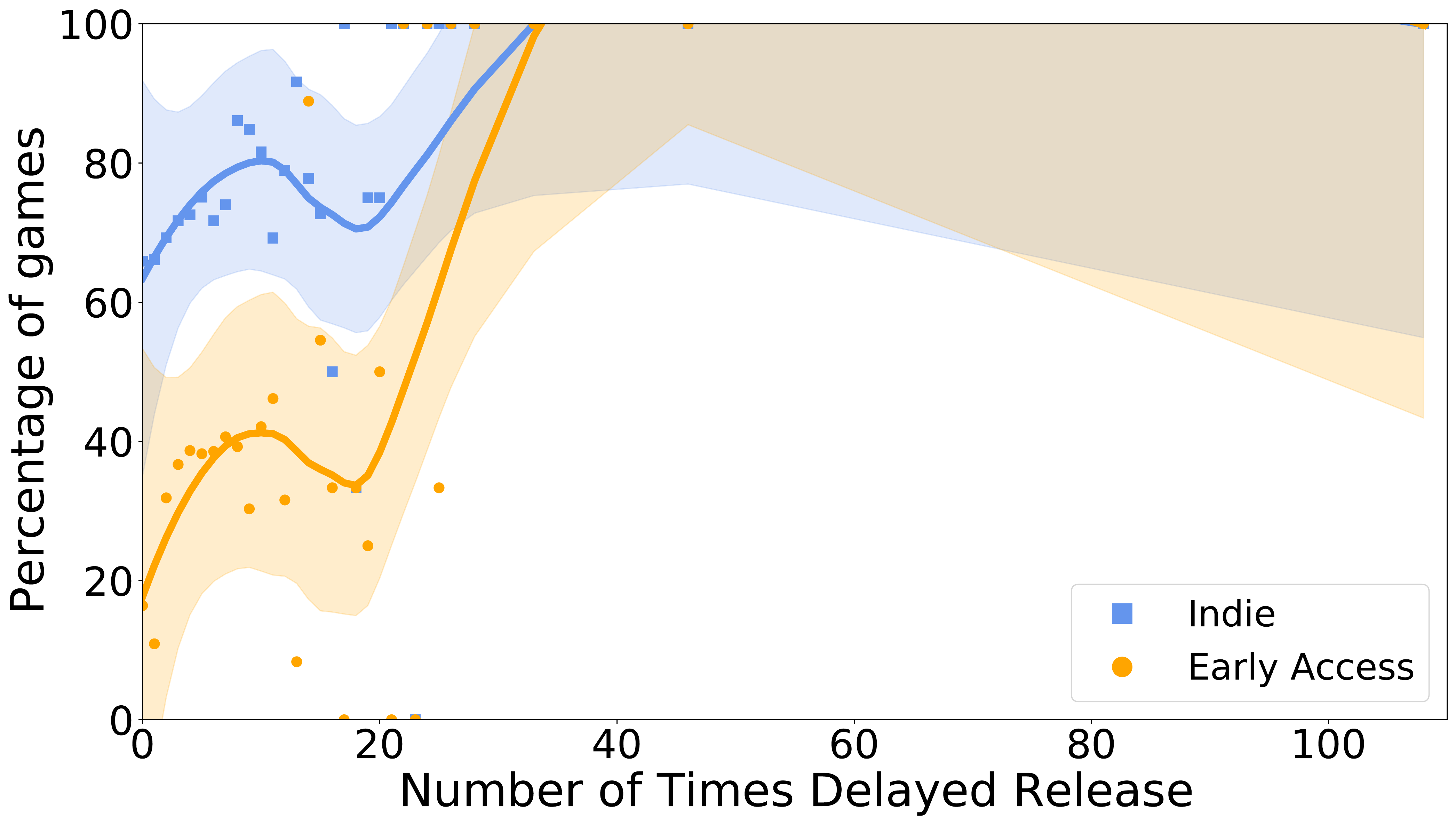}
\caption{An overview of the percentage of early access and indie games vs. the number of times a game was delayed. The trend lines are LOESS trend lines which provide a smoothed estimate of the percentage of games with the 95\% confidence interval. Note that the confidence interval is very wide for games that were delayed more than 30 times since we do not have many of such games in our data set.}
\label{fig:EA}
\end{figure}

\textit{Findings:}
\textbf{As the number of times a game is delayed increased, the percentage of early access games and indie games also increased.} As seen in Figure~\ref{fig:EA}, 11\% of the games that are delayed once were early access games. However among games that were delayed more often, the percentage of early access games increased. Early access games delayed a median of two times, one more time than the median value for all games as shown in Section~\ref{section:RQ1}.

Similar to early access games, as the number of times a game delays release increased, the portion of indie games also increased. Initially, as seen in Figure~\ref{fig:EA}, indie games that delayed release once account for 66\% of all games. The percentage of indie games remains stable ($\pm$5\%) until the number of times delayed surpasses 7 from where the percentage of indie games increases and then decreases followed by a sharp increase. 

\textbf{The percentage of games that delayed release across different genres was similar.} Compared to the 48\% of all games with a concrete release date that delayed release, 48\% of adventure and action games delayed release. The percentage of RPG and casual games that delayed release was 49\%. The games with the largest percentage of delayed releases were strategy (51\%) and simulation (52\%) games. However, these percentages were fairly close to the 48\% of all games (regardless of their genre) that delayed release suggesting that the genre of a game does not determine if the game is likely to delay its release.

\begin{tcolorbox}[left=2pt, right=2pt, top=2pt, bottom=2pt]
  \textit{\textbf{Summary:} As the number of times a game delayed its release increased, the percentage of games that were early access or indie also increased. The genre of a game did not determine if a game was likely to delay its release.}
\end{tcolorbox}

\section{\Qthree}
\label{section:RQ3}

\textit{Motivation:} User reviews give an indication of how well a game is perceived by players~\cite{Lin2018reviews}. For developers, the most important review metric is the positive review rate, as games with a low positive review rate are unlikely to be successful~\cite{Lin16eag}. Many different factors can be correlated with the positive review rate of a game. Hence, in this RQ, we study if overall, delayed games tend to have different positive review rates than on-time games.

\begin{figure}[t!]
\centering
\includegraphics[width=0.8\textwidth]{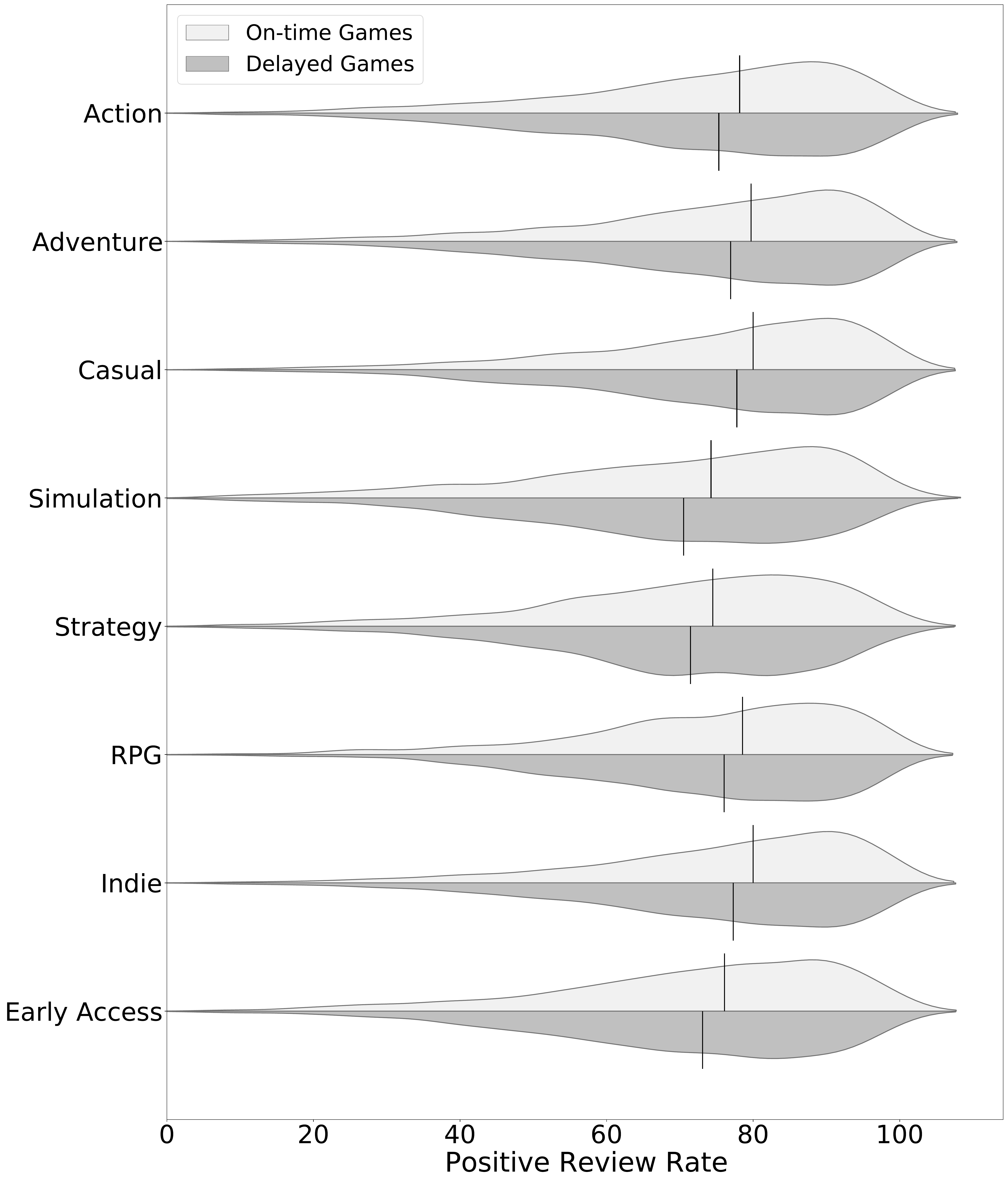}
\caption{Beanplots of the positive review rates of the studied types of delayed and on-time games. The vertical bar displays the median positive review rate for each distribution. }
\label{fig:PosRev}
\end{figure}

\textit{Approach:} For this research question, we used a custom crawler to extract the number of reviews, and the number of positive reviews from the Steam platform for each unique AppID. On Steam, users rate a game as `Recommended' or `Not Recommended'. Using the number of `Recommended' ratings for a game, we calculated the positive review ratings for games with at least 10 reviews using the following formula as used on Steam~\cite{Lin2018reviews, SteamDB}: $\textit{Positive review rate} = \frac{\textit{\# of recommended reviews}}{\textit{\# of all reviews}}\mathit{*100}$. To ensure accurately measuring the differences between delayed and on-time games, we removed games with a non-concrete release date for this RQ.

We compared the distributions of delayed and on-time games for the game types that we specified in Section~\ref{section:method} using the Wilcoxon rank-sum test~\cite{Wilcoxon}. The Wilcoxon rank-sum test is a non-parametric statistical test which is used to compare two ordinal distributions. The null hypothesis is that the two distributions being compared are similar and can be rejected at a p-value of less than 0.05, suggesting that the two distributions are in fact significantly different. To understand the magnitude of the difference between two ordinal distributions, we utilize Cliff's delta, $d$. We use thresholds for $d$ as presented by Romano et al.~\cite{romano2006exploring}:

\begin{equation*}
    \text{effect size} = \left\{
    \begin{array}{@{}ll@{}}
        \text{negligible} & \text{if } |d| \leq 0.147 \\
        \text{small} & \text{if } 0.147 < |d| \leq 0.33 \\
        \text{medium} & \text{if } 0.33 < |d| \leq 0.474 \\
        \text{large} & \text{if } 0.474 < |d| \leq 1 \\
    \end{array}
\right.
\end{equation*}

\begin{figure}[t!]
\centering
\includegraphics[width=0.8\textwidth]{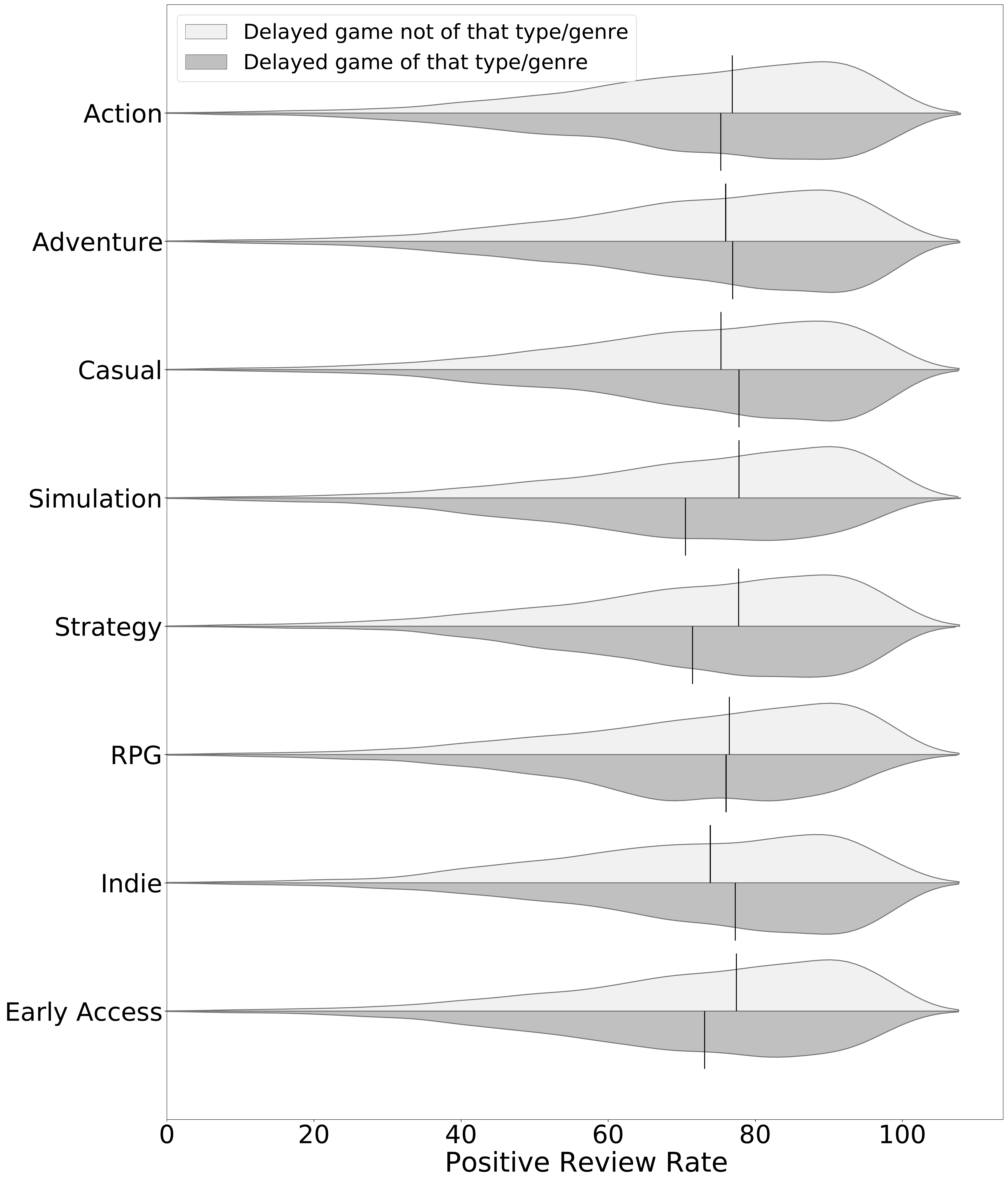}
\caption{Beanplots of the positive review rates of the studied types of games that delayed release compared to all other games. The vertical bar displays the median positive review rate for each distribution. }
\label{fig:PosRevTag}
\end{figure}

Because we conducted six comparisons, significant results could appear due to chance. Hence, we applied the Bonferroni correction to $\alpha$, resulting in $\alpha = 0.0083$ for Figure~\ref{fig:PosRevTag}.

\textit{Findings:}
\textbf{The median values of the positive review rates of games were slightly different within each studied genre and type when comparing delayed and on-time games.} The positive review rate of various game genres and types that delayed release or were on-time are presented in Figure~\ref{fig:PosRev}. A series of Wilcoxon rank-sum tests for comparing the distributions of delayed and on-time games for each genre suggest that all of the distributions are significantly different with negligible effect sizes ($|d|=0.06$ for action and casual games, $|d|=0.07$ adventure $|d|=0.08$ for simulation and indie games, and $|d|=0.06$ for strategy and RPG games) with the exception of early access games for which the distributions are not significantly different.

\textbf{Delayed casual and indie games had a greater positive review rate than other delayed games, and delayed simulation and strategy games had a lower positive review rate than other delayed games.} The positive review rates of games of various game genres and types that delayed, compared with all games not of that genre or type that delayed are shown in Figure~\ref{fig:PosRevTag}. For example, delayed action games are compared with delayed non-action games in the top beanplot in the figure. Wilcoxon rank-sum tests for the distributions for each game type or genre and the effect sizes (with the Bonferroni correction applied) are presented in Table~\ref{table:PosRev}. Delayed casual and indie games were perceived better than other delayed games as the median values for the positive review rate was greater for these games than others, however the effect sizes of the distributions were negligible (see Table~\ref{table:PosRev}). On the other hand, delayed simulation games had a lower positive review rate than other delayed games with a small effect size. Delayed strategy games also had a lower positive review rate than other delayed games, however, the effect size was negligible. 

\begin{table*}[t]
    \centering
    \caption{Wilcoxon rank-sum tests for comparing distributions of the positive review rates of delayed games within the type/genre compared to all other games outside of the type/genre}
    \label{table:PosRev}
    \begin{tabular}{l r r}
    \toprule
    \textbf{Game genre} & \textbf{Distributions} & \textbf{Effect size ($|d|$)} \\
    \midrule
    Action & Similar & -\\
    Adventure & Similar & -\\
    Casual & Different & 0.05 (negligible)\\
    Simulation & Different & 0.19 (small)\\
    Strategy & Different & 0.15 (negligible)\\
    RPG & Similar & -\\
    \midrule
    \textbf{Game type} & \textbf{Distributions} & \textbf{Effect Size ($|d|$)}\\
    \midrule
    Indie & Different & 0.07 (negligible)\\
    Early access & Different & 0.17 (small)\\
    \bottomrule
    \end{tabular}
\end{table*}

\begin{tcolorbox}[left=2pt, right=2pt, top=2pt, bottom=2pt]
  \textit{\textbf{Summary:} Games that delayed release had a slightly lower positive review rate than games that were released on-time.}
\end{tcolorbox}

\section{\Qfour}
\label{section:RQ4}

\textit{Motivation:} 
If developers are unsure of exactly when a game will be ready to release, there is the option to specify a non-concrete date rather than an exact date (e.g., ``Q1 2020"). These non-concrete dates  give developers the flexibility to specify a date range initially and choose the exact release date at a later time. Thus, here we analyze games with a non-concrete release date to measure how likely it is for such games to delay their release. An understanding of the accuracy of non-concrete release dates could assist developers when choosing release dates for their games.

\begin{table*}[t]
\centering
\caption{Non-concrete release dates}
\label{table:non-concrete}
\begin{tabular}{l l r}
 \toprule
 \textbf{Type} & \textbf{Example} & \textbf{Number of Games}\\
 \midrule
 Year & `2016' & 636\\
 Quarter + year & `Q3 2018' & 549\\
 Month + year & `May 2020' & 98\\
 Season & `Winter' & 65\\
TBA/TBC/TBD & `TBA' & 129\\
 Soon & `soon' & 979\\
 Other & `when ready' & 1,608\\
 \bottomrule
\end{tabular}
\end{table*}

\textit{Approach:} 
We extracted the 4,064 games that initially have a non-concrete release date, then categorized the games based on the type of non-concrete release date using regular expressions. We distinguish the following types of non-concrete release dates: year only, the quarter and year of release, and the month and year of release. We also examine ambiguous non-concrete release dates, such as those specified as ``soon", ``TBA"/``TBC"/``TBD", or a season of release. 

After categorizing the games based on their non-concrete release date types, we counted the number of games that meet the date ranges by matching the non-concrete release date to the actual release date. For example, if a game specified a non-concrete release date of ``Q1 2019" and released on February 3, 2019, we would first ensure that the year of release and the year of the non-concrete release are the same, then we would ensure that the quarter matched the month of the release (in this case February is in Quarter 1). As long as the specified dates in the non-concrete release matched the final release date, we considered the game to meet its non-concrete release date. Note, that we did not count the number of times a game with a non-concrete release was delayed.

We also counted the number of days until a game enters its non-concrete release date range on Steam. Since there is no initial concrete release date in the release date history for games with a non-concrete release date, we counted the number of days until a game enters its non-concrete release date range by calculating the number of days between the announcement of the non-concrete release date and the starting date of the non-concrete release date of the game. For example, if a game sets a non-concrete release date of ``July 2020" on May 2, 2020, we would count the number of days between May 2, 2020 and July 1, 2020 (this being the start of the release date range) to calculate the number of days until the release (60 days in this case). We set games that posted their non-concrete release date while within the release date range to have entered the release date range in 0 days. 

To compare the characteristics of the number of days until a non-concrete release dates enters its specified release date range we used the Wilcoxon rank-sum test and we measured the effect size using Cliff's delta, $d$, as mentioned previously in Section~\ref{section:RQ3}. Because we conducted three comparisons, significant results could appear due to chance. Hence, we applied the Bonferroni correction to $\alpha$, resulting in $\alpha = 0.0167$.

To calculate the number of days until a game with a non-concrete release date is released, we calculated the number of days from the announcement of the non-concrete release date until the actual release date. For example, if a game sets a non-concrete release date of ``Q2 2019" on March 3, 2019 and then the game releases September 13, 2019, we counted the number of days between the two dates (meaning the game released in 194 days).

To compare the distributions for the number of days until the release of different non-concrete release date types we use the Scott-Knott clustering algorithm using a 95\% significance level. The Scott-Knott clustering algorithm ranks the distributions of values. Distributions that are similar are placed in the same rank.  

\begin{figure}[t]
\centering
\includegraphics[width=0.8\textwidth]{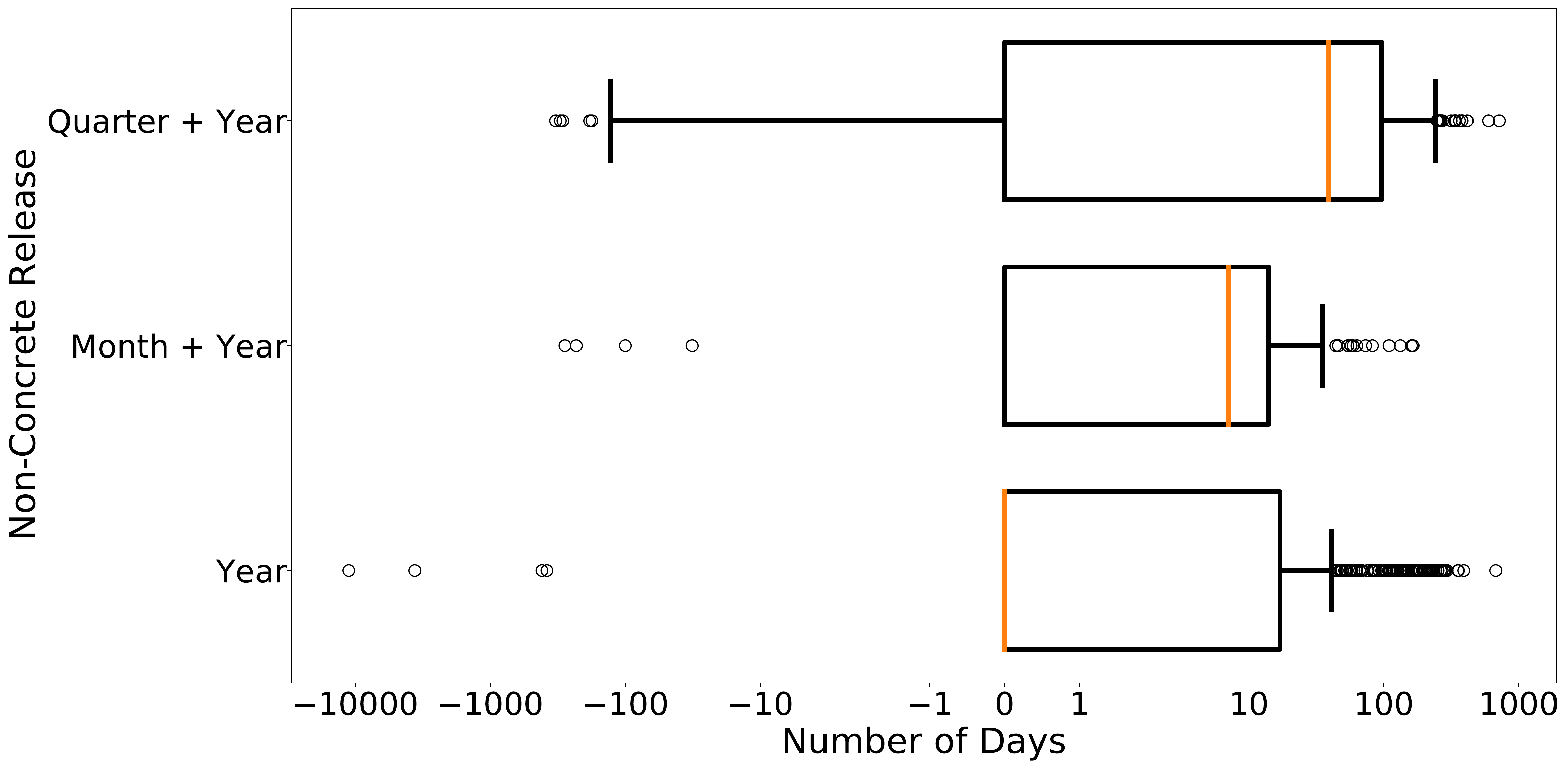}
\caption{Boxplots of the number of days a non-concrete release date is announced before entering the non-concrete release date range. Note that games that release in 0 days are games that post the non-concrete release date range while in it. The negative numbers of days indicate a placeholder release date range.}
\label{fig:NonConcRel}
\end{figure}

\textit{Findings:} 
\textbf{Non-concrete release dates were more likely to be met than concrete release dates, unless they were of the quarter and year type.} Compared to the 52\% of games with concrete release dates that were on-time, 67\% of games that gave only a year of release met their non-concrete release date. Similarly, 69\% of games that gave the month and year of release met the range. Of the games that specified the quarter and year of release only 44\% met the specified range, lower than games with a concrete release date. Hence, in most cases, non-concrete release dates appear to offer developers enough room to adjust the final release date of a game without having to post multiple different release dates over the span of the game's development period. As we showed in Section~\ref{section:RQ1}, the median number of days by which a game is delayed is 14. Hence, choosing a non-concrete release date allows developers to cover that period, helping them to make it more likely that the release date is made.

Interestingly, our observation holds for most types of non-concrete release dates, except for the quarter and year type. A possible explanation is that developers who choose to specify a release date of this type are not very clear yet about the actual release date of their games. 

\textbf{Games that give a month and year of release, or a quarter and year of release, posted their non-concrete release dates before entering the release date range set by the non-concrete release date.} The distributions of the number of days from a game posting a non-concrete release date to entering the non-concrete release date range can be seen in Figure~\ref{fig:NonConcRel}. Games that specify a quarter and year of release post the initial non-concrete release date a median of 39 days before the non-concrete release date range begins. Games that post only a year of release, do so a median of 0 days (meaning while in the non-concrete release date range) until the non-concrete release date range begins. These games meet their non-concrete release dates more often because of the large range a year encompasses for releasing and thus developers have more flexibility. Games that are of the month and year type post a release date a median of 7 days before entering the intended month of release.

The Wilcoxon rank-sum test shows that the distributions in Figure~\ref{fig:NonConcRel} are significantly different, with small (year compared to year and month with $|d|=0.19$) or medium effect sizes (year compared to quarter and year with $|d|=0.37$, and quarter and year compared to year and month with $|d|=0.33$).

\begin{figure}[t]
\centering
\includegraphics[width=0.8\textwidth]{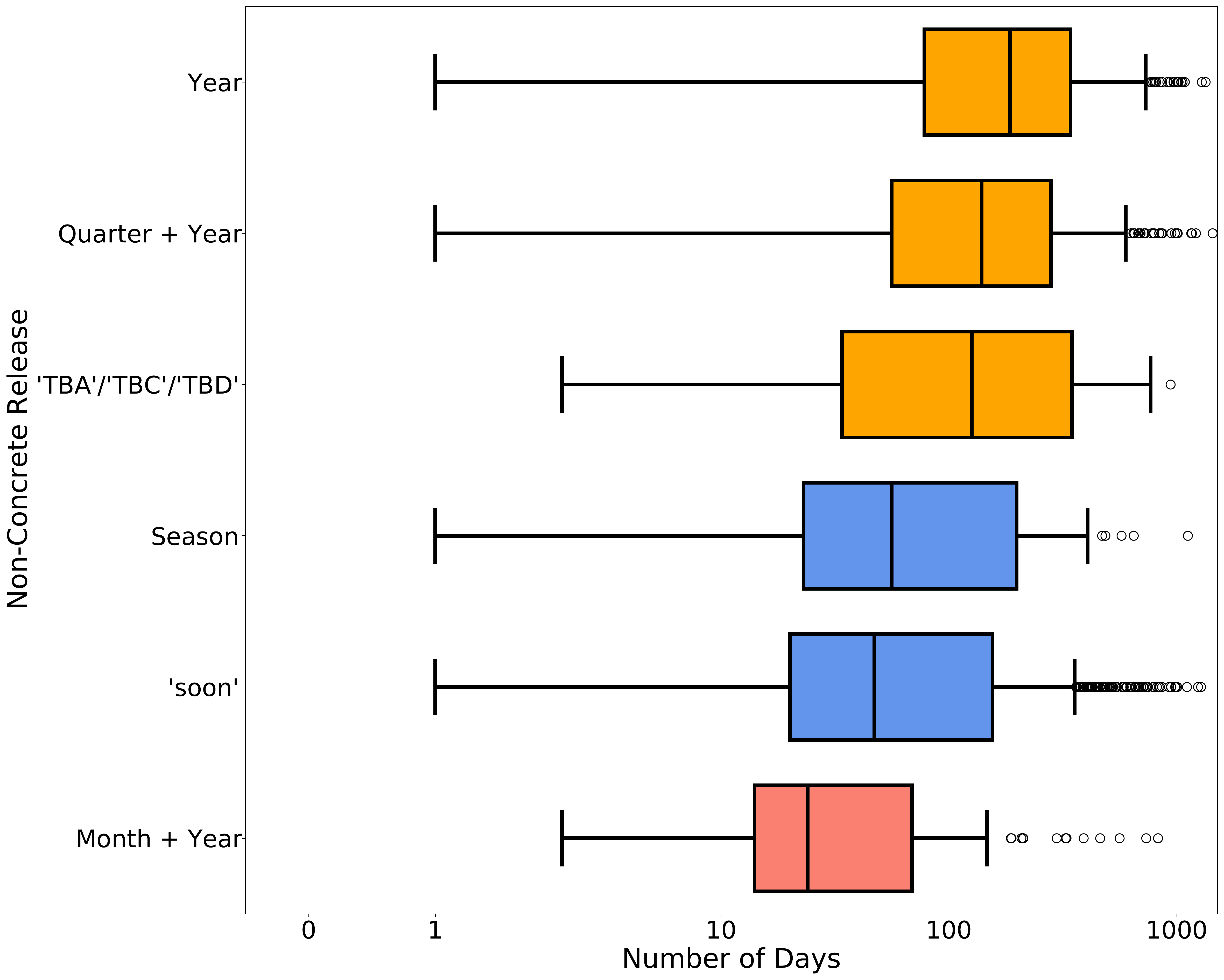}
\caption{Boxplots of the number of days between posting a non-concrete release date and releasing the game. Each colour exhibits a rank according to the Scott-Knott test; non-concrete release date types in different ranks have significantly different distributions.}
\label{fig:Ambigious}
\end{figure}

\textbf{Most games with ambiguous non-concrete release dates release within 3 months while games that specify a year of release or quarter and year of release took longer.} Non-concrete release dates that do not have a particular release date range are not bound to release in any amount of time, however, as seen in Figure~\ref{fig:Ambigious}, most games with ambiguous release dates released within 3 months. Games that specified they would release ``soon" were released within a median of 47 days. Similarly, games that specified a season of release (i.e., Winter, Summer, Fall/Autumn, Spring), released within a median of 56 days. On the other hand, games that specified ``TBA''/``TBC''/``TBD'' as their release date, released within a median of 126 days. Games with a release date range as a non-concrete release date take longer than 3 months to release as seen in Figure~\ref{fig:Ambigious}, with the exception of games that specify a release month and year which release within a median of 24 days. Games that specify the year of release take a median of 186 days to release, and games that specify a quarter and year of release within a median of 139 days.

Interestingly, games that specify the quarter and year of release, actually release past the 90 day length of a quarter. Taking into account the median of 39 days that games of the quarter and year type post their non-concrete release date before entering the specified release period, the games still release a median of 10 days following the end of the quarter specified. The possible explanation for this behaviour may be that developers are not clear about when their game will release when specifying a quarter of release.

The Scott-Knott clustering algorithm placed games that specify a year of release, a quarter and year of release, and ``TBA"/``TBC"/``TBD" in the same rank, suggesting that the distributions are similar. The algorithm next placed games that specify the keyword ``soon" and a season of release in the same rank. Games that specify the month and year of release were ranked separately suggesting that games of this type are significantly different from the the other types of games. This may be due to the small timeline for release a month encompasses compared to the longer releases of the other release date ranges.

\begin{tcolorbox}[left=2pt, right=2pt, top=2pt, bottom=2pt]
  \textit{\textbf{Summary:} Non-concrete release dates(games with no definitive release date such as ``May 2022”) were more likely to be met than concrete release dates (such as ``02/05/2022”) as long as the non-concrete release date did not consist of the year and quarter of release. Games with a non-concrete release date consisting of only the year posted its release date while in the release date range. Other types of games with a non-concrete release date posted their release dates prior to entering the release date range. Games with ambiguous non-concrete release dates(such as ``soon”) released within three months. Games that specify a year of release or a quarter and year of release took longer than 3 months to release.}
\end{tcolorbox}

\section{Discussion of our findings}
\label{section:Discuss}

Our findings indicate that delaying a release is becoming an increasingly common practice in game development. Over time more than 50\% of games have delayed their release date at least once to allow for slightly more flexibility in the development process. 

Furthermore, we found evidence that different genres and game types delaying release had only a slightly lower positive review rate than games of other genres and game types that delay release. Thus, a developer of a certain style of game does not necessarily need to be wary about delaying the release of their game when it comes to the positive review rate. This finding can aid developers in being more comfortable changing their release date to allow for flexibility in their game development cycle. 

If developers are unsure about when they plan to release their game, we find that non-concrete release dates can be a good option. These release dates allow for a developer to estimate a release time and give a more general date than committing to a specific date in the early steps of development. Giving a non-concrete release date gives flexibility for developers to change the final release date based on the current status of the game while still meeting the originally planned release projections. 

Although our findings suggest that choosing a release date can be done strategically, it should be acknowledged that there are a multitude of external factors that can affect the final release date and perceived quality of a game. However, as explained above, our data shows that there are several factors related to choosing a release date that are correlated with the game being released on time.

\section{Threats to Validity}
\label{section:threat}

\textit{Internal validity:} A potential threat to the validity of our study is that we only analyze data before the COVID-19 pandemic. However, this choice was on purpose as we could not collect enough data yet to give a reliable analysis of changes during the pandemic. Games released or set to release during the pandemic may have been subject to a change in trends. Future studies should repeat our analysis on data after April 2020 to understand how trends have changed during the COVID-19 pandemic. 

Another threat to the validity of our findings is that we only studied non-concrete release dates that were written in English. Future studies should investigate if our findings hold for non-concrete release dates in other languages.

A threat to the validity of the positive review rate, used in Section~\ref{section:RQ3}, are review bombs. Review bombs occur when a large number of negative reviews are posted on a game to discredit it. Though review bombs are uncommon, they can negatively impact the positive review rate of a game. Steam does aim to review and remove review bombs if a developer has opted for it \cite{revBomb}, but we encourage future studies to further filter games to reveal the impact of review bombs.

Another threat to the validity of our study is the classification of indie games. We find that the definition of an Indie game is not precisely defined. Thus, we use the genre tags provided by the developer to classify a game as indie. This may result in our findings not generalizing to games on other platforms, as the definition and tagging of indie games may vary.

A final threat to validity is that we cannot verify if a game was in fact released on the Steam platform and not abandoned by developers. We assume that games with at least one review were released on Steam  (as only games that are released can have a review). 94\% of the games analyzed in this study have at least one review, hence, the effect of games with no reviews should be negligible. Future studies should implement a process to verify if the games analyzed are available on Steam and were not abandoned by developers. 

\textit{External validity:}  In our study we only examine PC games on the Steam platform. Our findings may not generalize to games on other technical and distribution platforms. For example, the rating of a game is calculated differently by gaming platforms depending on the method of reviews used by the platform; thus a game can potentially have different ratings across platforms. 

Further, a threat to the validity of our findings is the games that prior to 2019 abused the ability to delay their release in order to stay on top of certain pages on Steam. These games have a very large value for the number of times released. However, we chose to not filter these games out and leave them as outliers in our data because the percentage of these types of games are very small ($<$ 1\%). Future studies should implement a process to verify that the games analyzed did not abuse the ability to delay their release.

\section{Conclusion}
\label{section:conclude}

In this paper, we studied 9,262 delayed and 10,159 on-time games. In particular, we studied how often different types of games were delayed, how players perceived delayed games, and how often non-concrete release dates were met. Our most notable findings are:
\begin{enumerate}
    \item 48\% of the games delayed their release, and the portion of games that delay is increasing.
    \item Non-concrete release dates are more likely to be met by games, unless these dates are of the quarter and year type.
    \item Games with a delayed release are rated lower than games that release on time, but the difference is negligible.
    \item The percentage of games that delay release across different game genres is similar (48\% to 52\%).
\end{enumerate}

Our findings show that delaying the release of a game is a common practice in game development. Hence, if it is necessary to delay an initial release date to ensure the quality of the final product, developers could take our study into account when deciding if they should delay a release. 

While our findings do not consider the financial outcomes of delaying the release date, we found little evidence of a relationship between delaying release and positive review rates. Thus, the received perception of a game is probably more attributed to its quality rather than release date.

Future studies should explore the trends of delaying release during the COVID-19 pandemic and more closely analyze user reviews to understand if there are complaints regarding delayed releases. 

\bibliographystyle{elsarticle-harv}
\bibliography{main}

\begin{thebibliography}{38}
\expandafter\ifx\csname natexlab\endcsname\relax\def\natexlab#1{#1}\fi
\providecommand{\url}[1]{\texttt{#1}}
\providecommand{\href}[2]{#2}
\providecommand{\path}[1]{#1}
\providecommand{\DOIprefix}{doi:}
\providecommand{\ArXivprefix}{arXiv:}
\providecommand{\URLprefix}{URL: }
\providecommand{\Pubmedprefix}{pmid:}
\providecommand{\doi}[1]{\href{http://dx.doi.org/#1}{\path{#1}}}
\providecommand{\Pubmed}[1]{\href{pmid:#1}{\path{#1}}}
\providecommand{\bibinfo}[2]{#2}
\ifx\xfnm\relax \def\xfnm[#1]{\unskip,\space#1}\fi
\bibitem[{Bailey and Miyata(2019)}]{BAILEY2019100299}
\bibinfo{author}{Bailey, E.}, \bibinfo{author}{Miyata, K.},
  \bibinfo{year}{2019}.
\newblock \bibinfo{title}{Improving video game project scope decisions with
  data: An analysis of achievements and game completion rates}.
\newblock \bibinfo{journal}{Entertainment Computing} \bibinfo{volume}{31},
  \bibinfo{pages}{100299}.
\bibitem[{Blackburn et~al.(2014)Blackburn, Kourtellis, Skvoretz, Ripeanu and
  Iamnitchi}]{BlackburnCheatingGames}
\bibinfo{author}{Blackburn, J.}, \bibinfo{author}{Kourtellis, N.},
  \bibinfo{author}{Skvoretz, J.}, \bibinfo{author}{Ripeanu, M.},
  \bibinfo{author}{Iamnitchi, A.}, \bibinfo{year}{2014}.
\newblock \bibinfo{title}{Cheating in online games: A social network
  perspective}.
\newblock \bibinfo{journal}{ACM Trans. Internet Technol.} \bibinfo{volume}{13}.
\bibitem[{Chambers et~al.(2010)Chambers, Feng, Sahu, Saha and
  Brandt}]{ChambersOnlineGames}
\bibinfo{author}{Chambers, C.}, \bibinfo{author}{Feng, W.c.},
  \bibinfo{author}{Sahu, S.}, \bibinfo{author}{Saha, D.},
  \bibinfo{author}{Brandt, D.}, \bibinfo{year}{2010}.
\newblock \bibinfo{title}{Characterizing online games}.
\newblock \bibinfo{journal}{IEEE/ACM Transactions on Networking}
  \bibinfo{volume}{18}, \bibinfo{pages}{899--910}.
\bibitem[{Einav(2010)}]{EinavRelTiming}
\bibinfo{author}{Einav, L.}, \bibinfo{year}{2010}.
\newblock \bibinfo{title}{Not all rivals look alike: Estimating an equilibrium
  model of the release date timing game}.
\newblock \bibinfo{journal}{Economic Inquiry} \bibinfo{volume}{48},
  \bibinfo{pages}{369--390}.
\bibitem[{Einav and Ravid(2009)}]{Einav2009}
\bibinfo{author}{Einav, L.}, \bibinfo{author}{Ravid, S.A.},
  \bibinfo{year}{2009}.
\newblock \bibinfo{title}{Stock market response to changes in movies' opening
  dates}.
\newblock \bibinfo{journal}{Journal of Cultural Economics}
  \bibinfo{volume}{33}, \bibinfo{pages}{311--319}.
\bibitem[{Engelst{\"a}tter and Ward(2018)}]{Engel2018}
\bibinfo{author}{Engelst{\"a}tter, B.}, \bibinfo{author}{Ward, M.R.},
  \bibinfo{year}{2018}.
\newblock \bibinfo{title}{Strategic timing of entry: evidence from video
  games}.
\newblock \bibinfo{journal}{Journal of Cultural Economics}
  \bibinfo{volume}{42}, \bibinfo{pages}{1--22}.
\bibitem[{Epp et~al.(2021)Epp, Lin and Bezemer}]{rain2021vr}
\bibinfo{author}{Epp, R.}, \bibinfo{author}{Lin, D.}, \bibinfo{author}{Bezemer,
  C.P.}, \bibinfo{year}{2021}.
\newblock \bibinfo{title}{An empirical study of trends of popular virtual
  reality games and their complaints}.
\newblock \bibinfo{journal}{IEEE Transactions on Games} ,
  \bibinfo{pages}{1--12}.
\bibitem[{Foxman et~al.(2020)Foxman, Leith, Beyea, Klebig, Chen and
  Ratan}]{FoxmanVR}
\bibinfo{author}{Foxman, M.}, \bibinfo{author}{Leith, A.P.},
  \bibinfo{author}{Beyea, D.}, \bibinfo{author}{Klebig, B.},
  \bibinfo{author}{Chen, V.H.H.}, \bibinfo{author}{Ratan, R.},
  \bibinfo{year}{2020}.
\newblock \bibinfo{title}{Virtual reality genres: Comparing preferences in
  immersive experiences and games}, in: \bibinfo{booktitle}{Extended Abstracts
  of the 2020 Annual Symposium on Computer-Human Interaction in Play},
  \bibinfo{publisher}{Association for Computing Machinery},
  \bibinfo{address}{New York, NY, USA}. p. \bibinfo{pages}{237–241}.
\bibitem[{Fu et~al.(2013)Fu, Lin, Li, Faloutsos, Hong and Sadeh}]{FuWisCom}
\bibinfo{author}{Fu, B.}, \bibinfo{author}{Lin, J.}, \bibinfo{author}{Li, L.},
  \bibinfo{author}{Faloutsos, C.}, \bibinfo{author}{Hong, J.},
  \bibinfo{author}{Sadeh, N.}, \bibinfo{year}{2013}.
\newblock \bibinfo{title}{Why people hate your app: Making sense of user
  feedback in a mobile app store}, in: \bibinfo{booktitle}{Proceedings of the
  19th ACM SIGKKD International Conference on Knowledge Discovery and Data
  Mining}, \bibinfo{publisher}{Association for Computing Machinery},
  \bibinfo{address}{New York, NY, USA}. p. \bibinfo{pages}{1276–1284}.
\bibitem[{Galyonkin(2021)}]{Steamspy}
\bibinfo{author}{Galyonkin, S.}, \bibinfo{year}{2021}.
\newblock \bibinfo{title}{Steamspy - all the data and stats about {S}team
  games}.
\newblock \URLprefix \url{https://steamspy.com}.
\bibitem[{Hassan et~al.(2019)Hassan, Bezemer and Hassan}]{safwat_tse}
\bibinfo{author}{Hassan, S.}, \bibinfo{author}{Bezemer, C.P.},
  \bibinfo{author}{Hassan, A.E.}, \bibinfo{year}{2019}.
\newblock \bibinfo{title}{Studying bad updates of top free-to-download apps in
  the google play store}.
\newblock \bibinfo{journal}{The Transactions of Software Engineering (TSE)
  journal} .
\bibitem[{Hassan et~al.(2018)Hassan, Tantithamthavorn, Bezemer and
  Hassan}]{safwat16replies}
\bibinfo{author}{Hassan, S.}, \bibinfo{author}{Tantithamthavorn, C.},
  \bibinfo{author}{Bezemer, C.P.}, \bibinfo{author}{Hassan, A.E.},
  \bibinfo{year}{2018}.
\newblock \bibinfo{title}{Studying the dialogue between users and developers of
  free apps in the {G}oogle {P}lay store}.
\newblock \bibinfo{journal}{The Empirical Software Engineering (EMSE) journal}
  \bibinfo{volume}{23}, \bibinfo{pages}{1275--1312}.
\bibitem[{Hu et~al.(2018)Hu, Wang, Bezemer and Hassan}]{hu17hybrid}
\bibinfo{author}{Hu, H.}, \bibinfo{author}{Wang, S.}, \bibinfo{author}{Bezemer,
  C.P.}, \bibinfo{author}{Hassan, A.E.}, \bibinfo{year}{2018}.
\newblock \bibinfo{title}{Studying the consistency of star ratings, reviews and
  releases of top free hybrid {A}ndroid and {iOS} apps}.
\newblock \bibinfo{journal}{The Empirical Software Engineering (EMSE) journal}
  \bibinfo{volume}{24}, \bibinfo{pages}{7--32}.
\bibitem[{Humphries(2019)}]{PCSteamRel}
\bibinfo{author}{Humphries, M.}, \bibinfo{year}{2019}.
\newblock \bibinfo{title}{Steam release date changes now require valve's
  approval}.
\newblock \bibinfo{howpublished}{PCMag}.
\newblock \URLprefix
  \url{https://www.pcmag.com/news/steam-release-date-changes-now-require-valves-approval}.
\bibitem[{Jeffrey D.~Long(2003)}]{Wilcoxon}
\bibinfo{author}{Jeffrey D.~Long, Du~Feng, N.C.}, \bibinfo{year}{2003}.
\newblock \bibinfo{title}{Ordinal analysis of behavioral data}.
\newblock \bibinfo{journal}{Handbook of psychology} ,
  \bibinfo{pages}{635–661}.
\bibitem[{Lee et~al.(2019)Lee, Lin, Bezemer and Hassan}]{Daniel2019nexusmods}
\bibinfo{author}{Lee, D.}, \bibinfo{author}{Lin, D.}, \bibinfo{author}{Bezemer,
  C.P.}, \bibinfo{author}{Hassan, A.E.}, \bibinfo{year}{2019}.
\newblock \bibinfo{title}{Building the perfect game - an empirical study of
  game modifications}.
\newblock \bibinfo{journal}{Empirical Software Engineering} .
\bibitem[{Lee et~al.(2020)Lee, Rajbahadur, Lin, Sayagh, Bezemer and
  Hassan}]{Lee2020curseforge}
\bibinfo{author}{Lee, D.}, \bibinfo{author}{Rajbahadur, G.K.},
  \bibinfo{author}{Lin, D.}, \bibinfo{author}{Sayagh, M.},
  \bibinfo{author}{Bezemer, C.P.}, \bibinfo{author}{Hassan, A.E.},
  \bibinfo{year}{2020}.
\newblock \bibinfo{title}{An empirical study of the characteristics of popular
  minecraft mods}.
\newblock \bibinfo{journal}{Empirical Software Engineering Journal} .
\bibitem[{Li and Zhang(2020)}]{LiGameTags}
\bibinfo{author}{Li, X.}, \bibinfo{author}{Zhang, B.}, \bibinfo{year}{2020}.
\newblock \bibinfo{title}{A preliminary network analysis on steam game tags:
  Another way of understanding game genres}, in:
  \bibinfo{booktitle}{Proceedings of the 23rd International Conference on
  Academic Mindtrek}, \bibinfo{publisher}{Association for Computing Machinery},
  \bibinfo{address}{New York, NY, USA}. p. \bibinfo{pages}{65–73}.
\bibitem[{Lin et~al.(2017)Lin, Bezemer and Hassan}]{Lin16urgent}
\bibinfo{author}{Lin, D.}, \bibinfo{author}{Bezemer, C.P.},
  \bibinfo{author}{Hassan, A.E.}, \bibinfo{year}{2017}.
\newblock \bibinfo{title}{Studying the urgent updates of popular games on the
  {S}team platform}.
\newblock \bibinfo{journal}{The Empirical Software Engineering (EMSE) journal}
  \bibinfo{volume}{22}, \bibinfo{pages}{2095--2126}.
\bibitem[{Lin et~al.(2018a)Lin, Bezemer and Hassan}]{Lin16eag}
\bibinfo{author}{Lin, D.}, \bibinfo{author}{Bezemer, C.P.},
  \bibinfo{author}{Hassan, A.E.}, \bibinfo{year}{2018}a.
\newblock \bibinfo{title}{An empirical study of early access games on the
  {S}team platform}.
\newblock \bibinfo{journal}{The Empirical Software Engineering (EMSE) journal}
  \bibinfo{volume}{23}, \bibinfo{pages}{771--799}.
\bibitem[{Lin et~al.(2019)Lin, Bezemer and Hassan}]{Lin2019videos}
\bibinfo{author}{Lin, D.}, \bibinfo{author}{Bezemer, C.P.},
  \bibinfo{author}{Hassan, A.E.}, \bibinfo{year}{2019}.
\newblock \bibinfo{title}{Identifying gameplay videos that exhibit bugs in
  computer games}.
\newblock \bibinfo{journal}{Empirical Software Engineering} .
\bibitem[{Lin et~al.(2018b)Lin, Bezemer, Zou and Hassan}]{Lin2018reviews}
\bibinfo{author}{Lin, D.}, \bibinfo{author}{Bezemer, C.P.},
  \bibinfo{author}{Zou, Y.}, \bibinfo{author}{Hassan, A.E.},
  \bibinfo{year}{2018}b.
\newblock \bibinfo{title}{An empirical study of game reviews on the steam
  platform}.
\newblock \bibinfo{journal}{Empirical Software Engineering} .
\bibitem[{Nolibois(2021)}]{accenture}
\bibinfo{author}{Nolibois, Q.}, \bibinfo{year}{2021}.
\newblock \bibinfo{title}{Global gaming industry value now exceeds \$300
  billion, new accenture report finds}.
\newblock \bibinfo{howpublished}{Newsroom}.
\newblock \URLprefix
  \url{https://newsroom.accenture.com/news/global-gaming-industry-value-now-exceeds-300-billion-new-accenture-report-finds.htm}.
\bibitem[{Pagano and Maalej(2013)}]{PagnoApp}
\bibinfo{author}{Pagano, D.}, \bibinfo{author}{Maalej, W.},
  \bibinfo{year}{2013}.
\newblock \bibinfo{title}{User feedback in the appstore: An empirical study},
  in: \bibinfo{booktitle}{2013 21st IEEE International Requirements Engineering
  Conference (RE)}, pp. \bibinfo{pages}{125--134}.
\bibitem[{Romano et~al.(2006)Romano, Kromrey, Coraggio, Skowronek and
  Devine}]{romano2006exploring}
\bibinfo{author}{Romano, J.}, \bibinfo{author}{Kromrey, J.D.},
  \bibinfo{author}{Coraggio, J.}, \bibinfo{author}{Skowronek, J.},
  \bibinfo{author}{Devine, L.}, \bibinfo{year}{2006}.
\newblock \bibinfo{title}{Exploring methods for evaluating group differences on
  the {NSSE} and other surveys: Are the t-test and {C}ohen’s d indices the
  most appropriate choices}, in: \bibinfo{booktitle}{annual meeting of the
  Southern Association for Institutional Research},
  \bibinfo{organization}{Citeseer}. pp. \bibinfo{pages}{1--51}.
\bibitem[{Sifa et~al.(2014)Sifa, Bauckhage and Drachen}]{SifaCrossGames}
\bibinfo{author}{Sifa, R.}, \bibinfo{author}{Bauckhage, C.},
  \bibinfo{author}{Drachen, A.}, \bibinfo{year}{2014}.
\newblock \bibinfo{title}{The playtime principle: Large-scale cross-games
  interest modeling}, in: \bibinfo{booktitle}{2014 IEEE Conference on
  Computational Intelligence and Games}, pp. \bibinfo{pages}{1--8}.
\bibitem[{{Steamworks development}(2019)}]{SteamRelChange}
\bibinfo{author}{{Steamworks development}}, \bibinfo{year}{2019}.
\newblock \bibinfo{title}{Release date tool changes}.
\newblock \bibinfo{howpublished}{Steam Community}.
\newblock \URLprefix
  \url{https://steamcommunity.com/groups/steamworks/announcements/detail/1595883107395323259}.
\bibitem[{Tong(2021)}]{TONG2021100077}
\bibinfo{author}{Tong, X.}, \bibinfo{year}{2021}.
\newblock \bibinfo{title}{Positioning game review as a crucial element of game
  user feedback in the ongoing development of independent video games}.
\newblock \bibinfo{journal}{Computers in Human Behavior Reports}
  \bibinfo{volume}{3}, \bibinfo{pages}{100077}.
\bibitem[{{Valve Corporation}(2019)}]{revBomb}
\bibinfo{author}{{Valve Corporation}}, \bibinfo{year}{2019}.
\newblock \bibinfo{title}{User reviews revisited}.
\newblock \bibinfo{howpublished}{Steam Community}.
\newblock \URLprefix
  \url{https://steamcommunity.com/games/593110/announcements/detail/1808664240333155775?snr=2_groupannouncements_detail_}.
\bibitem[{{Valve Corporation}(2021a)}]{SteamCommunity}
\bibinfo{author}{{Valve Corporation}}, \bibinfo{year}{2021}a.
\newblock \bibinfo{title}{Steam {C}ommunity}.
\newblock \URLprefix \url{https://steamcommunity.com}.
\bibitem[{{Valve Corporation}(2021b)}]{SteamStore}
\bibinfo{author}{{Valve Corporation}}, \bibinfo{year}{2021}b.
\newblock \bibinfo{title}{Steam {S}tore}.
\newblock \URLprefix \url{https://store.steampowered.com}.
\bibitem[{{Valve Corporation}(2021c)}]{Steamworks}
\bibinfo{author}{{Valve Corporation}}, \bibinfo{year}{2021}c.
\newblock \bibinfo{title}{Steamworks}.
\newblock \URLprefix \url{https://partner.steamgames.com}.
\bibitem[{Vasa et~al.(2012)Vasa, Hoon, Mouzakis and Noguchi}]{VasaMobileApp}
\bibinfo{author}{Vasa, R.}, \bibinfo{author}{Hoon, L.},
  \bibinfo{author}{Mouzakis, K.}, \bibinfo{author}{Noguchi, A.},
  \bibinfo{year}{2012}.
\newblock \bibinfo{title}{A preliminary analysis of mobile app user reviews},
  in: \bibinfo{booktitle}{Proceedings of the 24th Australian Computer-Human
  Interaction Conference}, \bibinfo{publisher}{Association for Computing
  Machinery}, \bibinfo{address}{New York, NY, USA}. p.
  \bibinfo{pages}{241–244}.
\bibitem[{Vu and Bezemer(2021)}]{Quang21}
\bibinfo{author}{Vu, Q.N.}, \bibinfo{author}{Bezemer, C.P.},
  \bibinfo{year}{2021}.
\newblock \bibinfo{title}{Improving the discoverability of indie games by
  leveraging their similarity to top-selling games identifying important
  requirements of a recommender system}, in: \bibinfo{booktitle}{International
  Conference on the Foundations of Digital Games (FDG)}, pp.
  \bibinfo{pages}{1--12}.
\bibitem[{Wang and Goh(2020)}]{WANG2020100338}
\bibinfo{author}{Wang, X.}, \bibinfo{author}{Goh, D.H.L.},
  \bibinfo{year}{2020}.
\newblock \bibinfo{title}{Components of game experience: An automatic text
  analysis of online reviews}.
\newblock \bibinfo{journal}{Entertainment Computing} \bibinfo{volume}{33},
  \bibinfo{pages}{100338}.
\bibitem[{{xP}aw and Marlamin(2021)}]{SteamDB}
\bibinfo{author}{{xP}aw}, \bibinfo{author}{Marlamin}, \bibinfo{year}{2021}.
\newblock \bibinfo{title}{Steam {D}atabase - {S}team{DB}}.
\newblock \URLprefix \url{https://steamdb.info}.
\bibitem[{Yannakakis(2012)}]{YannakakisAI}
\bibinfo{author}{Yannakakis, G.N.}, \bibinfo{year}{2012}.
\newblock \bibinfo{title}{Game {AI} revisited}, in:
  \bibinfo{booktitle}{Proceedings of the 9th Conference on Computing
  Frontiers}, \bibinfo{publisher}{Association for Computing Machinery},
  \bibinfo{address}{New York, NY, USA}. p. \bibinfo{pages}{285–292}.
\bibitem[{Zhu and Fang(2015)}]{ZhuGamePlay}
\bibinfo{author}{Zhu, M.}, \bibinfo{author}{Fang, X.}, \bibinfo{year}{2015}.
\newblock \bibinfo{title}{A lexical approach to study computer games and game
  play experience via online reviews}.
\newblock \bibinfo{journal}{International Journal of Human–Computer
  Interaction} \bibinfo{volume}{31}, \bibinfo{pages}{413--426}.

\end{thebibliography}

\end{document}